\documentclass[a4paper,11pt]{article}
\usepackage{amssymb}
\usepackage{hyperref}
\title{The Saeed-Liu-Tian-Gao-Li authenticated key agreement protocol is insecure}
\author{Chris J. Mitchell\\Information Security Group, Royal Holloway, University of London\\
\url{www.chrismitchell.net}}
\date{21st June 2019}
\begin{document}

\bibliographystyle{plain}

\maketitle

\section*{Abstract}

A recently proposed authenticated key agreement protocol is shown to be
insecure. In particular, one of the two parties is not authenticated, allowing
an active `man in the middle' opponent to replay old messages.  The protocol is
essentially an authenticated Diffie-Hellman key agreement scheme, and the lack
of authentication allows an attacker to replay `old' messages and have them
accepted. Moreover, if the ephemeral key used to compute a protocol message is
ever compromised, then the key established using the replayed message will also
be compromised. Fixing the problem is simple --- there are many provably secure
and standardised protocols which are just as efficient as the flawed scheme.

\section{Introduction} \label{Intro}

In a recent paper, Saeed et al.\ \cite{Saeed19} proposed an authenticated key
agreement (AKA) protocol, based on an El Gamal-type identity-based signature
scheme and Diffie-Hellman key agreement in an elliptic curve setting. The
scheme is intended for use in the Internet of Things (IoT), but there appears
to be little about the protocol that makes it uniquely suitable for this
domain.

In this brief paper we first outline the protocol and then demonstrate that it
fails to achieve authentication of either party.  That is, an active attacker
can successfully impersonate either party in the protocol at any time, as long
as the attacker first successfully intercepts the messages sent in one genuine
instance of the protocol. Moreover, if an ephemeral key ever becomes available
to an attacker, then the attacker can perform the protocol with a genuine party
and convince that party to use a session key known to the attacker. These
observations break (at least) two of the claimed properties of the protocol.

\section{Outline of the AKA protocol}

As outlined in Section~\ref{Intro}, the Saeed et al.\ protocol \cite{Saeed19}
is claimed to provide `authenticated key agreement'; it seems reasonable to
assume that this, amongst other things, captures the notion that both parties
are given assurance that the other party is involved in the protocol, i.e.\
that both parties are `live'.  However, it is hard to identify any part of the
security model and security definition which captures this notion; this is
rather surprising, and the authors appear to have departed from the usual
security notions for such protocols. Perhaps not surprisingly, as we discuss
below, the protocol fails to achieve proof of liveness.

The authors claim a list of security properties for the protocol
--- however, these properties are not all supported by the claimed security
proof in the paper'.  Two of these properties are as follows.
\begin{itemize}
\item {\bf P4}: Key compromise impersonate;
\item {\bf P5}: Replay attack.
\end{itemize}
As we discuss below, neither of these properties hold for the protocol.  That
is, we show that the protocol fails to achieve two properties specifically
claimed for it.

The protocol involves two messages exchanged between a server and a client
(referred to in \cite{Saeed19} as a `sensor node').  The server sends a message
to the client and receives a response in return.  The two messages have exactly
the same structure and content types, and the second message appears to be
completely independent of the first.  That is, it would appear that they could
be exchanged in either order or, indeed, in parallel.

All cryptographic computations involve a pre-agreed elliptic curve $\mathbb{C}$
defined over a finite field $\mathbb{F}$, together with a pre-agreed point $P$
on this curve that generates a cyclic (additive) group $\mathbb{G}$ on the
points of the curve.  The system uses identity-based signatures, and thus there
is a universally trusted private key generator (PKG) who has a master
private/pubic key pair ($s_{\mathrm{PKG}}$, $S_{\mathrm{PKG}}$), where
$s_{\mathrm{PKG}}$ is an integer ($0<s_{\mathrm{PKG}}<q$) and
$S_{\mathrm{PKG}}=s_{\mathrm{PKG}}P$. Every entity who participates in the
protocol (client or server) is assumed to have a well-defined identifier ID,
and is pre-equipped by the PKG with a long-term private/public key pair ($s$,
$R$), where $R=rP$ for a randomly chosen integer $r$; $s=(r+cs_{\mathrm{PKG}})P
\bmod q$; and $c=H_1(\mbox{ID}||R)$ for a suitable hash function $H_1$.  Since
$s_{\mathrm{PKG}}$ is known only to the PKG, only the PKG can generate key
pairs.

Before constructing a protocol message, the sender generates two ephemeral
(one-time) values $X$ and $Y$.  The ephemeral public key $Y$ is simply a scalar
multiple $yP$ of the point $P$, where the integer $y$ ($0<y<q$) is a randomly
chosen ephemeral private key.  Similarly, the value $X$ is a scalar multiple
$xP$ of the point $P$, where the integer $x$ ($0<y<q$) is also randomly chosen
by the sender ($x$ and $X$ form part of signature generation/verification).

The first message, i.e.\ that sent from the server to the client, has the
following form:

\[ \mbox{ID} || Y || \sigma || t \]

where ID is an identifier for the sender (in this case the server), $Y$ is an
ephemeral public key created by the sender, $\sigma$ is a type of El Gamal
signature, $t$ is a (fresh) timestamp created by the sender, and, as
throughout, $||$ denotes concatenation of data items (possibly incorporating
encoding to ensure unambiguous decoding by the recipient). The second message
is identical to the first except that in each case the sender is the client.

The signature $\sigma$ is a triple ($h$, $\mu$, $R$), computed as follows.
\begin{itemize}
\item $h=H_2(\mbox{ID}|\mbox{ID}'||Y||R||X)$, where $H_2$ is an appropriate
    hash function, $\mbox{ID}$ is the identifier of the sender,
    $\mbox{ID}'$ is the identifier of the intended recipient, $Y$ is the
    ephemeral public key of the sender, $R$ is the long-term public key of
    the sender, and $X$ is an ephemeral value generated by the sender ---
    see above);
\item $\mu=x+hs$, where $x$ is an ephemeral value generated by the sender
    and $s$ is the long-term private key of the sender; and
\item $R$ is the long-term public key of the sender.
\end{itemize}

The message recipient (the client, although exactly the same process applies
for the client-server message) performs the following steps to verify the
message and establish a shared secret session key.
\begin{enumerate}
\item The freshness of the timestamp $t$ in the message is checked.
\item Compute $c=H_1(\mbox{ID}||R)$, where ID and $R$ are the values in the
    received message, and
\[  h^*=H_2(\mbox{ID}||\mbox{ID}'||Y||R||\mu P-h(R+cP_{\mathrm{PKG}})), \]
where ID is the identifier in the message, ID$'$ is the recipient's own
identifier, and $Y$, $R$, $\mu$ and $h$ are the values in the message, and
check that $h=h^*$;
\item Derive the shared secret key by hashing together the identifiers of
    the two parties (always server ID first) and the scalar multiple $y'Y$,
    where $y'$ is the recipient's own ephemeral private key.
\end{enumerate}

\section{A simple message replay attack}

It should be clear that the timestamp $t$ in the message is not verifiable in
any way, that is it is not incorporated into any of the cryptographically
computed parts of the message.  It could hence be modified by a
man-in-the-middle attacker without this change being detectable by the message
recipient.  This immediately suggests the following simple attack.

Suppose the attacker intercepts a message sent from the server to the client.
At any subsequent time the attacker can replay this message to the client and
it will be accepted as valid, as long as the attacker updates the timestamp $t$
before replaying the message. This immediately breaks claimed property {\bf P5}
(Replay attack resistance) of the protocol.  Note that a precisely analogous
attack enables impersonation of the client to the server.

Also, if the attacker compromises the ephemeral key $y$ corresponding to the
intercepted message sent from server to client, then the attacker can not only
impersonate the server at will, but can also calculate the session key for this
spurious session, since the only non-public input to the key calculation is the
ephemeral private key.  This shows that property {\bf P4} (Key compromise
impersonation resistance) also does not hold, contrary to what is claimed.

\section{Implications}

\subsection{Fixing the problem}

Of course, it would be relatively simple to fix the problem identified in this
short paper by including the value of the timestamp $t$ within the scope of
hash function $H_2$, ensuring that the signature $\sigma$ is a function of the
timestamp.  However, recommending such an ad hoc fix without more carefully
analysing the protocol would be unwise.

Moreover, this is an area in which there has been a great deal of research over
the last 20--30 years.  There are plenty of provably secure protocols
(including authenticated Diffie-Hellman key agreement schemes) which could be
used for the task addressed in this paper. In particular, the reader is
referred to ISO/IEC 11770-3 \cite{ISO11770-3:15}, which contains a wide variety
of authenticated key agreement protocols, all of which have been carefully
checked and validated. Also relevant is the book of Boyd and Mathuria
\cite{Boyd03} (and the new version to be published in late 2019), which
describes a very wide variety of such protocols.

\subsection{Other remarks}

There are a number of other serious issues with this paper, including the
following.
\begin{itemize}
\item Section 4.1 of the paper \cite{Saeed19} contains a proof of security.
    The fact that there are serious shortcomings in the protocol is
    therefore somewhat surprising.  However, as we have briefly observed
    above, the security model underlying the security result does not
    capture important properties.  In addition, on inspection, the proof of
    the main theorem lacks rigour, which means that there must be doubts
    about whether the theorem is true.
\item There are many problems with the exposition.  For example, there is
    confusion between the size of the underlying finite field and the
    number of points on the elliptic curve --- $q$ is used for both
    quantities.
\item Finally, although designed for lightweight sensors, the protocol
    requires all parties to have securely synchronised clocks and also the
    ability to generate high quality random numbers.  These are potentially
    unrealistic requirements for very lightweight devices.
\end{itemize}


\end{document}